\documentclass[twocolumn,twocolappendix]{aastex631}

\newcommand{\atob}{\text{ to }}
\newcommand{\keV}{\mathrm{keV}}
\newcommand{\Chandra}{\textit{Chandra}\ }
\newcommand{\WISE}{\textit{WISE}\ }
\newcommand{\Gaia}{\textit{Gaia}\ }

\newcommand{\kpc}{\mathrm{kpc}}
\newcommand{\Mpc}{\mathrm{Mpc}}
\newcommand{\logLx}{\log (L_\mathrm{X}/ \mathrm{erg\ s^{-1}})}

\newcommand{\Msunyr}{\mathrm{M_\odot \ yr^{-1}}}
\newcommand{\s}{\mathrm{s}}

\begin{document}

\title{Uncovering an Excess of X-ray Point Sources in the Halos of Virgo Late-type Galaxies}

\author[0009-0009-9972-0756]{Zhensong Hu}
\affiliation{School of Astronomy and Space Science, Nanjing University, Nanjing 210023, China}
\affiliation{Key Laboratory of Modern Astronomy and Astrophysics (Nanjing University), Ministry of Education, Nanjing 210023, China}
\email{huzhensong@smail.nju.edu.cn}

\author[0000-0001-9062-8309]{Meicun Hou}
\affiliation{Kavli Institute for Astronomy and Astrophysics, Peking University, Beijing 100871, China}
\email{houmc@pku.edu.cn}

\author[0000-0003-0355-6437]{Zhiyuan Li}
\affiliation{School of Astronomy and Space Science, Nanjing University, Nanjing 210023, China}
\affiliation{Key Laboratory of Modern Astronomy and Astrophysics (Nanjing University), Ministry of Education, Nanjing 210023, China}
\email{lizy@nju.edu.cn}

\begin{abstract}

We present a systematic search for extraplanar X-ray point sources around 19 late-type, highly inclined disk galaxies residing in the Virgo cluster, based on archival {\it Chandra} observations reaching a source detection sensitivity of $L\rm(0.5-8~keV)\sim10^{38}\rm~erg~s^{-1}$. 
Based on the cumulative source surface density distribution as a function of projected vertical distance from the disk mid-plane, 
we identify a statistically significant ($\sim3.3\sigma$) excess of $\sim20$ X-ray sources within a projected vertical off-disk distance of $0.92'-2.5'$ ($\sim$4.4--12 kpc),
the presence of which cannot be explained by the bulk stellar content of the individual galaxies, nor by the cosmic X-ray background.
On the other hand, there is no significant evidence for an excess of extraplanar X-ray sources in a comparison sample of field late-type edge-on galaxies, for which {\it Chandra} observations reaching a similar source detection sensitivity are available.
We discuss possible origins for the observed excess, which include low-mass X-ray binaries (LMXBs) associated with globular clusters, supernova-kicked LMXBs, 
high-mass X-ray binaries born in recent star formation induced by ram pressure stripping of the disk gas, as well as a class of intra-cluster X-ray sources previously identified around early-type member galaxies of Virgo.  
We find that none of these X-ray populations can naturally dominate the observed extraplanar excess, although supernova-kicked LMXBs and the effect of ram pressure are most likely relevant.
\end{abstract}


\section{Introduction} \label{sec:intro}

X-ray binaries (XRBs), in which a neutron or a stellar-mass black hole accretes from a companion star and shines at typical X-ray luminosities $\gtrsim10^{37}\rm~erg~s^{-1}$, are nowadays routinely detected as prominent point sources in nearby galaxies by \Chandra observations, thanks to its excellent angular resolution and sensitivity \citep{2002PASP..114....1W}. 
Broadly classified as low-mass X-ray binaries (LMXBs) and high-mass X-ray binaries (HMXBs) according to the mass of the companion, these XRBs can serve as a useful tool to reveal the nature of their parent stellar populations \citep{2006ARA&A..44..323F}. In particular, luminous and short-lived HMXBs are  abundant in late-type galaxies (LTGs) and closely related to recent and on-going star formation \citep{2003MNRAS.339..793G}, while LMXBs, having much longer evolution timescales, are ubiquitous in both late-type and early-type galaxies (ETGs) and are a good tracer of old stellar populations \citep[e.g.][]{2004MNRAS.349..146G,2004ApJ...611..846K}.

In recent years, a growing number of research suggest that a substantial fraction of X-ray sources also exist beyond the bulk stellar content of nearby ETGs.  For example, \citet{2010ApJ...721.1368L} detected $\sim100$ X-ray sources down to a limiting luminosity $L_{\rm X}\gtrsim10^{37}\ \mathrm{erg\ s^{-1}}$ in the halo of the Sombrero galaxy (M104), a field bulge-dominated Sa galaxy, and found a $\sim 4.4\sigma$ excess with respect to the expected number of sources arising from the cosmic X-ray background (CXB). They suggested that this excess is associated with the stellar halo of M104 and considered two possible origins: i) LMXBs formed in globular clusters (GC-LMXB), as a product of frequent stellar encounters within the dense GC environment, and ii) supernova-kicked (SN-kicked) LMXBs. 
\citet{2013A&A...556A...9Z} investigated the cumulative X-ray source surface density distribution of 20 nearby massive ETGs, 
from which they identified an overdensity of halo X-ray sources (also down to $L_{\rm X}\gtrsim 10^{37}\ \mathrm{erg\ s^{-1}}$) with respect to the combined contribution of LMXBs associated with the starlight (i.e., field LMXBs) and the CXB. These authors further estimated that SN-kicked LMXBs and GC-LMXBs account for $40\%$ and $60\%$ of the excess sources. 

More recent studies have provided evidence for an extended distribution of X-ray sources around ETGs residing in galaxy clusters, pointing to an environmental effect.
\citet{2017ApJ...846..126H} found a statistically significant ($3.5\sigma$) excess of X-ray sources at the outskirt of 80 Virgo low-to-intermediate mass ETGs when combining sources detected down to a luminosity limit of $L_{\rm X}\gtrsim2\times10^{38}\ \mathrm{erg\ s^{-1}}$. On the other hand, no significant excess of X-ray sources is found at the outskirt in their control sample of low-to-intermediate mass ETGs in the field. 
This led \citet{2017ApJ...846..126H} to propose the existence of intra-cluster X-ray sources (ICXs), which are currently gravitationally unbound to any member galaxy, analogous to and plausibly associated with the so-called diffuse intra-cluster light (ICL) conventionally detected in the optical/near-infrared band \citep{2022NatAs...6..308M}. 
The predominantly old stellar population of the ICL, as indicated by its red color \citep[e.g.][]{2017ApJ...834...16M}, is natural to harbor LMXBs (ICL-LMXBs). \citet{2017ApJ...846..126H} estimated that ICL-LMXBs may account for $\gtrsim10\%$ of the detected excess, 
while GC-LMXBs and SN-kicked LMXBs can also have a substantial contribution.
In a subsequent study focusing on the Fornax cluster, \citet{2019ApJ...876...53J} identified an excess with a significance of $3.6\sigma$ for X-ray sources (with $L_{\rm X}\gtrsim3\times10^{37}\ \mathrm{erg\ s^{-1}}$) detected outside three times the effective radius of the central giant elliptical galaxy NGC\,1399.  
The luminosity function of the excess sources was found to be steeper than that of the GC-LMXBs, implying that GC-LMXBs may have a minor contribution to the excess. Meanwhile, \citet{2019ApJ...876...53J} estimated that $\lesssim34\ (19\%)$ SN-kicked LMXBs are among the excess sources. Thus, these authors concluded that LMXBs associated with the extended stellar halo of NGC\,1399 and/or the ICL account for the majority of the excess sources.

Another environmental effect, which may contribute to the halo X-ray sources of a cluster member galaxy, is ram pressure, i.e., the interaction between the galaxy's interstellar medium (ISM) and the intra-cluster medium (ICM) \citep{2022A&ARv..30....3B}. Although ram pressure can strip gas from host galaxies and eventually halt star formation, it can also temporarily enhance star formation due to the increase in the pressure exerted on the disk of a galaxy \citep{2009A&A...499...87K,2012A&A...544A..54S}. 
Using the EAGLE simulation, \citet{2020MNRAS.497.4145T} find that the leading side (in the sense of motion inside the cluster) of cluster galaxies has enhanced ISM pressure and SFR. 
Such star forming activities boosted by ram pressure could result in a population of XRBs, in particular HMXBs, which, however, remain to be studied in detail.

The Virgo cluster, due to its proximity ($16.5\ \Mpc$; \citealp{2007ApJ...655..144M}) and abundant member galaxies ($\sim 1600$ identified;  \citealp{2014ApJS..215...22K}), is a unique laboratory for studying the environmental effect on the formation and evolution of various X-ray source populations outlined in the above, as already demonstrated by the work of \citet{2017ApJ...846..126H}. Moreover, a number of Virgo galaxies exhibits a disturbed X-ray morphology of the diffuse hot gas, including LTGs \citep[e.g.][]{2001A&A...380...40T,2004ApJ...610..183M,2011A&A...531A..44W}
and ETGs \citep[e.g.][]{2004ApJ...600..729R,2011ApJ...727...41K,2019AJ....158....6S,2021ApJ...919..141H},
which signifies ongoing ram pressure stripping. 
In this work, we conduct the first systematic search for excess X-ray sources in the halo of Virgo LTGs, as well as a sample of field LTGs for close comparison, for which sensitive {\it Chandra} observations are available.
Such a study, made possible by recently available {\it Chandra} observations, would complement the aforementioned studies of X-ray sources around ETGs and provide new insights into the evolution of XRBs in LTGs residing in the both the cluster and field environments.

This paper is organized as follows. In Section \ref{sec:data_preparation}, we describe sample selection and preparation of the X-ray data. 
In Section \ref{sec:analysis}, we present an analysis of the spatial distribution of X-ray sources in the Virgo LTGs, in a close comparison with a sample of field LTGs, which allows us to identify a statistically significant excess of off-disk sources in the former sample. Section~\ref{sec:discussion} 
addresses possible origins of this excess, with an emphasis on specific effects related to the cluster environment.
Finally, we provide a brief summary in Section \ref{sec:Summary}.

\section{Sample Selection and Data Preparation} \label{sec:data_preparation}

\subsection{Sample Selection}\label{subsec:SampleSelection}

We began with a parent sample of 75 Virgo cluster LTGs originally defined by  \citet{2022MNRAS.512.3284S}, which all have sufficiently deep {\it Chandra} observations for their primary goal of studying ultra-luminous X-ray sources and nuclear X-ray sources therein.
This sample spans a substantial range in galaxy morphology, including early-type spirals (i.e., with a substantial bulge), disk-dominated spirals and irregular galaxies. 
A recent work of \citet{2024ApJ...961..249H} selected among these 75 LTGs a subsample of 21 nearly edge-on disk galaxies to study the putative diffuse hot gas corona under the influence of the ICM. 
Since the focus of the present work is to probe an extraplanar population of discrete X-ray sources, we adopt the same nearly edge-on LTGs as in \citet{2024ApJ...961..249H}, with the exception of two galaxies, NGC\,4302 and NGC\,4343. NGC\,4302 is too close to its neighbor galaxy NGC\,4298, which makes it difficult to separate the X-ray sources belonging to the individual galaxies. 
For NGC\,4343, while it has two archival \Chandra observations, in both cases the bulk of the galaxy was placed at a relatively large off-axis angle, introducing an undesired strong asymmetry in the source detection sensitivity across the galaxy.
Therefore, we focus on 19 Virgo edge-on LTGs in this work (Table \ref{table:VL_property}). 
A visualization of the optical morphology of these galaxies, as well as their relative position in the Virgo cluster, can be found in \citet{2024ApJ...961..249H}.
We adopt a uniform distance of $16.5\ \Mpc$ \citep{2007ApJ...655..144M} for all of them.

To facilitate a direct comparison of the X-ray source distribution between the cluster and field environments,  
we built a comparison sample of nearly edge-on LTGs in the field. This field sample is again similar to the one used in \citet{2024ApJ...961..249H}.
Beginning from the highly inclined disk galaxy sample studied by \citet{2013MNRAS.428.2085L}, we filtered galaxies with the following criteria: (i) An inclination angle larger than $75^\circ$, to be consistent with the Virgo sample; (ii) A distance larger than $9\ \Mpc$ but smaller than $30\ \Mpc$, ensuring both a good coverage of the galactic disk/halo by the Chandra/ACIS and a good sensitivity for source detection; (iii) No large galaxy in the projected neighborhood. A total of twelve galaxies meet these criteria, with a median distance of $14.9\ \Mpc$ (Table \ref{table:VL_property}). 

\begin{figure*}
    \epsscale{1.2}
    \plotone{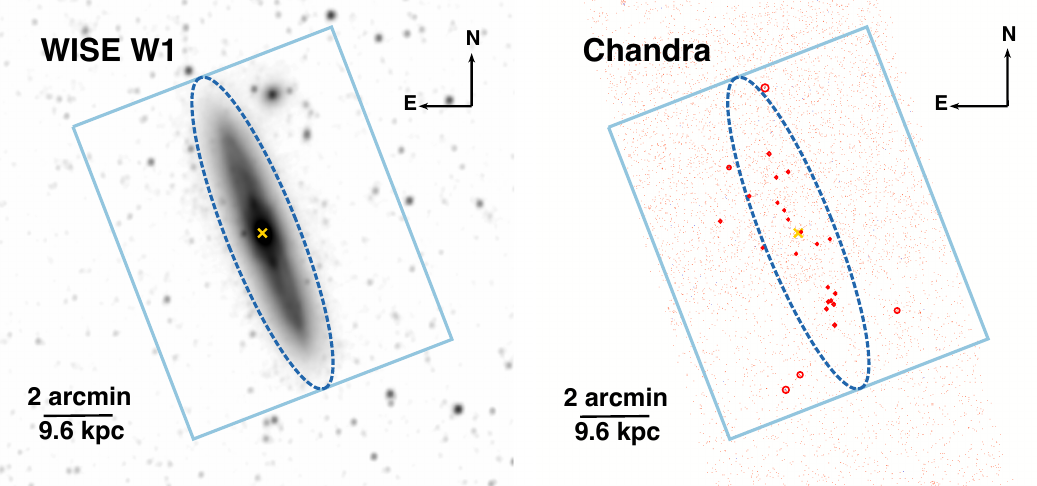}
    \caption{{\it Left}: The \WISE $W1$ image of NGC\,4216, a Virgo member galaxy, with a yellow ``$\times$" indicating the galactic center. The dark blue dashed ellipse shows the $22\ \mathrm{mag\ arcsec^{-2}}$ isophote of the galaxy, where its  major axis is defined as the disk mid-plane. The light blue solid box indicates the region within which the detected X-ray point sources are included for the spatial analysis. Its length is the same as the disk major axis, while its width is fixed to be $8'$. 
    {\it Right}: The {\it Chandra}/ACIS 0.5--8 keV counts image of NGC\,4216. The red circles marked the detected X-ray sources, with a radius of $2$ times the $90\%$ ECR. Other markers are the same as in the left panel.
    }
    \label{fig:NGC4216_WISE_Chandra}
\end{figure*}

To assist the analysis of the X-ray sources, we measured the galactic disk geometry, stellar mass and star formation rate (SFR) using the \WISE infrared images \citep{2010AJ....140.1868W}. The image of each galaxy was downloaded  from the WISE portal\footnote{https://irsa.ipac.caltech.edu/applications/wise/}. 
We define the galaxy size as the $22\ \mathrm{mag \ arcsec^{-2}}$ (Vega magnitude) isophote ($I_{22,W1}$) in the $W1$ band, which corresponds to the 2.5$\sigma$ sky surface brightness level \citep{2019ApJS..245...25J}. The isophotes were constructed with the Python tool {\tt photuitils}\footnote{https://photutils.readthedocs.io/en/stable/index.html} \citep{larry_bradley_2021_5796924}. 
The major-axis and position angle of the $I_{22,W1}$ isophote are taken to be the length and the position angle of the disk. 
The case of NGC\,4216 is shown as an example in Figure \ref{fig:NGC4216_WISE_Chandra}.
The stellar mass ($M_\star$) was then calculated using the \WISE stellar mass-luminosity ($L_{W1}$) relation \citep[Eq.~2 therein]{2019ApJS..245...25J}, $\log{M_\star/L_{W1} = -2.54(M_{W1}-M_{W2})-0.17}$, where $M_{W1}$ and $M_{W2}$ stand for the absolute magnitude of the target area (i.e., enclosed by $I_{22,W1}$) in the $W1$ ($3.4\ \mathrm{\mu m}$) and $W2$ ($4.6\ \mathrm{\mu m}$) images. $L_{W1}(L_{\odot,W1})=10^{-0.4(M-M_{\odot,W1})}$ is the corresponding luminosity at the $W1$ band and $M_{\odot,W1} = 3.24$ is the $W1$ band solar absolute magnitude. 
The median stellar mass of the Virgo and field sample is $\log({M_\star}/M_\odot) = 10.0$ and $\log({M_\star}/M_\odot) = 10.4$, respectively. 
The SFR was measured within the $I_{22,W1}$ isophote using the infrared SFR scaling relation of \citet[Eq.~3 therein]{2019ApJS..245...25J} based on the $W1$ and $W3$ images. 
The median SFR of the Virgo and field sample is $0.5\ \Msunyr$ and $2.8\ \Msunyr$, respectively.

In Table \ref{table:VL_property}, we provide the basic information for the sample galaxies, including their center coordinates, morphology type, size, stellar mass, SFR and distance.

\begin{deluxetable*}{ccccccccccccc}
\tablenum{1}
\tablecaption{Host Galaxy Information}
\tablewidth{0pt}
\tabletypesize{\footnotesize}
\tablehead{Galaxy & Morphology & Distance &  R.A. &   Decl. &  ObsID & Exp. & $N_{\rm src}$ &  log$M_*$ & SFR & $a$ & $b$ & PA  \\
           &    & (Mpc) &     (deg)  &    (deg)  &          & (ks)   &    & (M$_\odot$) & ($\Msunyr$)  &   (arcsec) & (arcsec) & (deg)   }
\decimalcolnumbers
\startdata
\multicolumn{13}{l}{Virgo} \\
\hline
NGC\,4192 & SABab & 16.5 & 183.451208 & 14.900333 & 19390 & 14.9 & 31 & 10.8 & 2.8 & 280 & 76 & 155 \\
NGC\,4197 & Sd & 16.5 & 183.660625 & 5.805722 & 19420 & 8.0 & 17 & 9.5 & 0.5 & 82 & 25 & 38 \\
NGC\,4206 & SAbc & 16.5 & 183.820042 & 13.023972 & 19431 & 8.0 & 15 & 10.0 & 0.3 & 142 & 29 & 0 \\
NGC\,4216 & SABb & 16.5 & 183.976833 & 13.149389 & 19391 & 9.5 & 26 & 11.0 & 1.7 & 290 & 62 & 22 \\
NGC\,4222 & Sd & 16.5 & 184.093833 & 13.307056 & 19432 & 8.0 & 10 & 9.4 & 0.3 & 111 & 21 & 58 \\
NGC\,4307 & Sb & 16.5 & 185.523667 & 9.043639 & 19405 & 14.7 & 24 & 10.1 & 0.4 & 119 & 38 & 27 \\
NGC\,4312 & SAab & 16.5 & 185.630667 & 15.537917 & 19414 & 8.0 & 17 & 10.1 & 0.7 & 125 & 44 & 171 \\
NGC\,4313 & SAab & 16.5 & 185.660583 & 11.800944 & 19416 & 8.0 & 10 & 10.1 & 0.4 & 135 & 45 & 138 \\
NGC\,4316 & Scd & 16.5 & 185.676000 & 9.332472 & 19400 & 14.9 & 18 & 9.7 & 0.6 & 76 & 31 & 115 \\
NGC\,4330 & Scd & 16.5 & 185.821875 & 11.367972 & 19419 & 8.0 & 9 & 9.8 & 0.4 & 140 & 26 & 56 \\
NGC\,4356 & Scd & 16.5 & 186.060542 & 8.535861 & 19435 & 14.9 & 21 & 10.0 & 0.1 & 98 & 33 & 40 \\
IC\,3322A & SBcd & 16.5 & 186.427333 & 7.216667 & 19401 & 13.9 & 22 & 9.5 & 0.5 & 104 & 22 & 156 \\
NGC\,4388 & SAb & 16.5 & 186.444792 & 12.662083 & 12291 & 27.6 & 29 & 10.0 & 2.5 & 176 & 57 & 91 \\
IC\,3322 & SABcb & 16.5 & 186.475417 & 7.554778 & 19424 & 14.9 & 19 & 9.3 & 0.2 & 72 & 21 & 158 \\
NGC\,4419 & SBa & 16.5 & 186.735167 & 15.047389 & 19394 & 9.9 & 18 & 10.3 & 1.6 & 102 & 51 & 134 \\
NGC\,4445 & Sab & 16.5 & 187.066375 & 9.436194 & 19433 & 14.9 & 23 & 9.7 & 0.2 & 78 & 27 & 103 \\
NGC\,4522 & SBcd & 16.5 & 188.415458 & 9.175028 & 19428 & 7.8 & 13 & 10.0 & 0.5 & 107 & 35 & 34 \\
NGC\,4532 & IBm & 16.5 & 188.580542 & 6.467694 & 19407 & 8.0 & 20 & 9.5 & 0.9 & 75 & 47 & 164 \\
NGC\,4607 & SBb & 16.5 & 190.301667 & 11.886639 & 19403 & 8.0 & 17 & 9.5 & 0.7 & 93 & 27 & 2 \\
\hline
\multicolumn{13}{l}{Field} \\
\hline
NGC\,24 & Sc & 9.1 & 2.485583 & -24.963139 & 9547 & 43.2 & 36 & 9.5 & 0.1 & 144 & 47 & 45 \\
NGC\,891 & Sb & 10.0 & 35.639224 & 42.349146 & 4613 & 118.9 & 119 & 10.4 & 5.5 & 349 & 93 & 24 \\
NGC\,3079 & SBcd & 16.5 & 150.490848 & 55.679789 & 19307 & 53.2 & 65 & 10.3 & 5.9 & 234 & 57 & 165 \\
NGC\,3556 & SBc & 10.7 & 167.879042 & 55.674111 & 2025 & 59.4 & 86 & 10.1 & 3.2 & 246 & 82 & 78 \\
NGC\,3628 & Sb & 13.1 & 170.070908 & 13.589489 & 2039 & 58.0 & 81 & 10.8 & 6.0 & 517 & 96 & 102 \\
NGC\,3877 & Sc & 14.1 & 176.532078 & 47.494346 & 1971 & 29.2 & 56 & 10.2 & 1.8 & 163 & 47 & 36 \\
NGC\,4013 & Sb & 18.9 & 179.630750 & 43.946583 & 4739 & 79.1 & 80 & 10.4 & 2.3 & 149 & 59 & 65 \\
NGC\,4217 & Sb & 19.5 & 183.962083 & 47.091778 & 4738 & 72.7 & 72 & 10.4 & 4.5 & 189 & 64 & 45 \\
NGC\,4565 & Sb & 11.1 & 189.086584 & 25.987675 & 3950 & 59.2 & 106 & 10.8 & 2.2 & 411 & 79 & 135 \\
NGC\,4666 & SABc & 15.7 & 191.285798 & -0.461885 & 21307 & 9.9 & 31 & 10.3 & 7.0 & 154 & 69 & 39 \\
NGC\,5170 & Sc & 22.5 & 202.453282 & -17.966419 & 3928 & 33.0 & 77 & 10.8 & 1.4 & 249 & 53 & 124 \\
NGC\,5746 & SABb & 24.7 & 221.232992 & 1.955003 & 3929 & 36.8 & 113 & 11.2 & 2.5 & 228 & 58 & 169 
\enddata 
\tablecomments{($ 1 $)  Name of the highly-inclined disk galaxies.  (2) Galaxy morphology type, from \citet{2022MNRAS.512.3284S} and \citet{2013MNRAS.428.2085L}. (3) Distances of the field galaxies, taken from \citet[Table 1 therein]{2013MNRAS.428.2085L}. The distance of all Virgo galaxies is set as $16.5\ \mathrm{Mpc}$ \citep{2007ApJ...655..144M}.  (4)-(5) Right ascension and declination at equinox J2000. (6)-(7) \Chandra observation ID and the corresponding exposure time.  (8) Number of detected X-ray sources.  (9) The total stellar mass measured with \textit{WISE} data, with an error of 0.1--0.2 dex. (10) The star formation rate, measured from the net flux in the \WISE $W3$ band by subtracting the stellar continuum derived from the $W1$ band, according to \citet[Eq.~3 therein]{2019ApJS..245...25J}. (11)-(13) The semi-major axis, semi-minor axis and position angle based on the measurement of the 22 $\mathrm{mag\ arcsec}^{-2}$ isophote in the \WISE $W1$ band.} 
\label{table:VL_property}
\end{deluxetable*}

\subsection{X-ray data processing} \label{subsec:datapreparation_X-ray}
\begin{figure*}[ht!]
    \epsscale{1.1}
    \plottwo{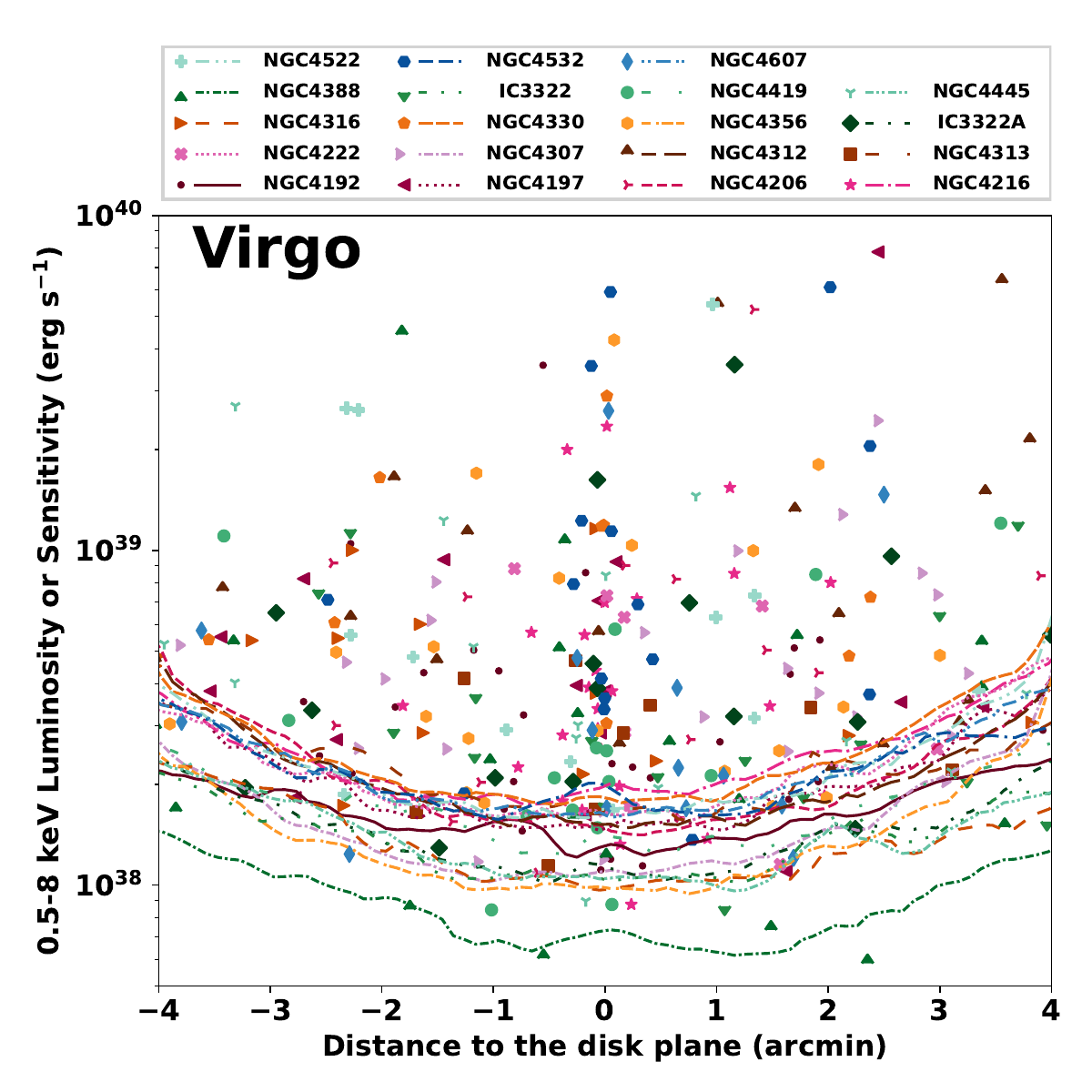}{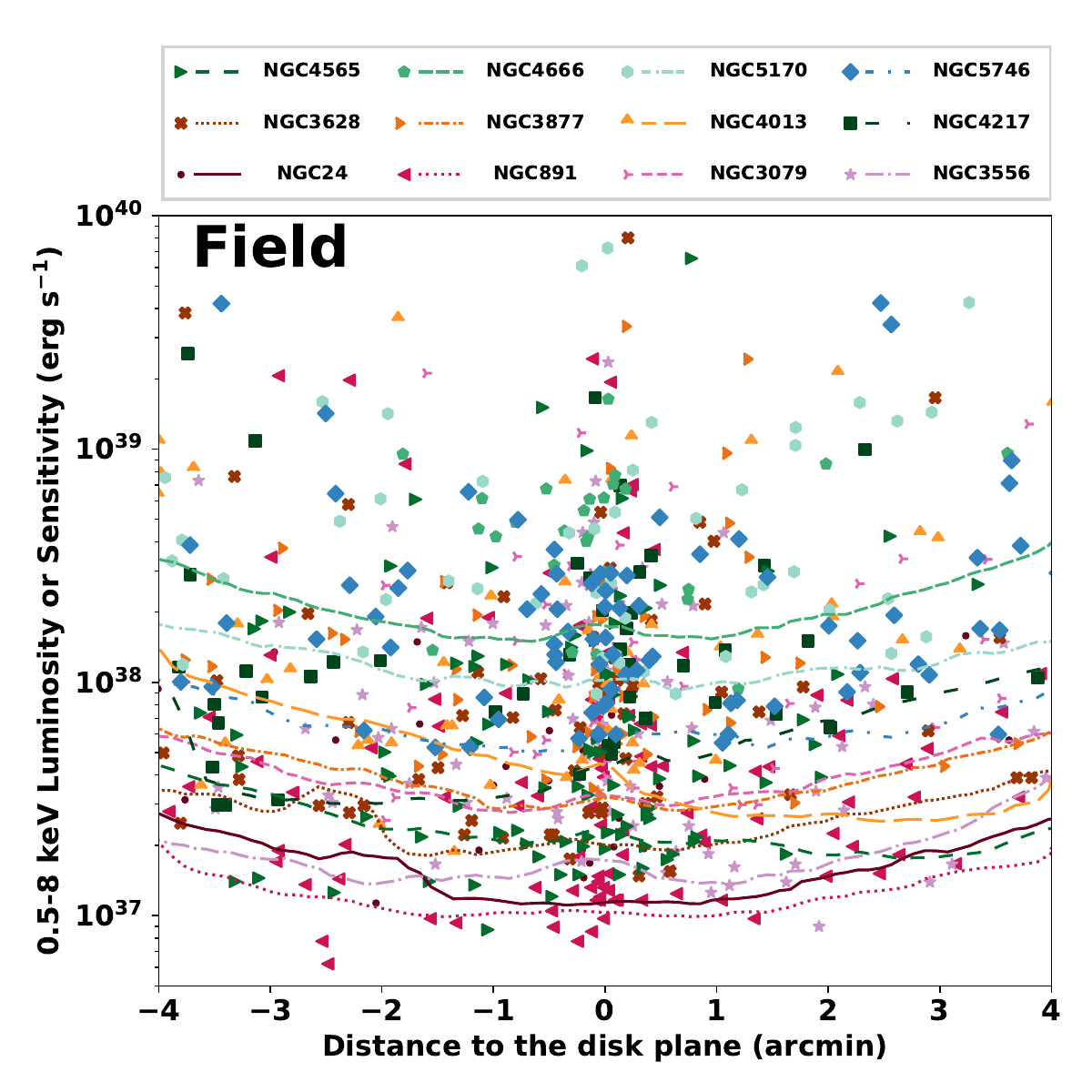}
    \caption{0.5--8 keV source luminosity (colored symbols) and median sensitivity limit (colored lines) vs. the vertical distance from the disk mid-plane, for the Virgo LTGs ({\it left panel}) and the field LTGs ({\it right panel}). 
    Identified AGNs and foreground stars are not included. 
    }
    \label{fig:Sensitivity}
\end{figure*}

All but one of the Virgo LTGs have only one {\it Chandra}/ACIS observation thus far, which is also the case for half of the field LTGs. For those target galaxies with more than one observations, we only used the one with the longest exposure, which suffices to reach the source detection sensitivity in those Virgo LTGs having a relatively short exposure (Table \ref{table:VL_property}; Figure~\ref{fig:Sensitivity}).

We downloaded and uniformly reprocessed the relevant {\it Chandra}/ACIS data following the standard procedures using CIAO v4.14 and the calibration files CALDB v4.9.7\footnote{http://cxc.harvard.edu/ciao/}.
We inspected the light curve of each observation and found that the instrumental background was quiescent for the vast majority of time intervals. Thus, all the science exposures were preserved for the subsequent analysis.
It occurs that in most observations the aimpoint was placed on the S3 CCD, for which we only used data from the S3 and S2 CCDs to ensure an optimal point-spread function (PSF). For the  observations of NGC\,5170 and NGC\,5746, the aimpoint was placed on the I3 CCD, so we only kept data from the I0, I1, I2 and I3 CCDs for the same reason.
For source detection and quantification, we focus on the 0.5--8 keV energy band, for which we generated a counts map from the level-2 events file.
To generate the exposure maps, we used the CIAO tool \texttt{mkexpmap}, assuming a fiducial incident power-law spectrum with a photon-index of $\Gamma=1.7$ and an absorption column density of $N_\mathrm{H}=10^{21}\ \mathrm{cm^{-2}}$, which is comparable to the Galactic foreground value of most sample galaxies.
Then we applied the CIAO tool \texttt{mkpsfmap} to produce the PSF map for a given enclosed count fraction (ECF). 
A point-source sensitivity map was generated based on the algorithm described by  \citet[]{2010ApJ...719..900K}.

We followed the procedures of \citet{2018ApJS..235...26Z} to detect and characterize discrete X-ray sources in each target galaxy of both the Virgo and field samples. First, we applied the CIAO tool \texttt{wavdetect} with the $50\%$-ECF PSF map in the $0.5-8\ \keV$ energy band. The false-positive detection threshold was set as $10^{-6}$.
Next, we refined the source centroid by iterating over the source counts inside the $90\%$ enclosed count radius (ECR). Thirdly, we performed source photometry by extracting source counts within the $90\%$ ECR. The background aperture was chosen as an annulus with inner-to-outer radii of 2--4 times the $90\%$ ECR. 
We calculated the binomial no-source probability $\mathcal{P}_B$ \citep{2007ApJ...657.1026W,2018ApJS..235...26Z}, defined as the probability of observing an equal or greater number of counts expected from the local background.  
Sources with a $\mathcal{P}_B$ value exceeding 0.01 were deemed spurious and subsequently excluded from the source list. 
We further checked all the sources by eye to avoid any spurious detection due to diffuse emission. No such case was found in the Virgo LTGs. 
However, for the field sample, we excluded several sources in NGC\,3079 and NGC\,3628 that appear to be associated with their bright central diffuse emission. 
Finally, the net photon flux was calculated with the CIAO tool \texttt{aprates} based on a Bayesian approach.

The above procedures resulted in a total of 359 sources for the nineteen Virgo LTGs and 922 sources for the twelve field LTGs.
The number of detected sources of each target field is presented in Table~\ref{table:VL_property}.
The higher abundance of X-ray sources in and around the field LTGs can be attributed to their on-average longer exposures, smaller distances, higher stellar masses and higher SFRs.

In Figure \ref{fig:Sensitivity}, we plot the unabsorbed 0.5--8 keV  luminosity of the detected sources as a function of the vertical distance ($z$) from the major-axis of the disk plane. We have adopted a photon flux to energy flux conversion factor of $3.29\times10^{-9}\ \mathrm{erg\ ph^{-1}}$, which is derived from the same fiducial incident power-law  for producing the exposure maps.  
For comparison, Figure \ref{fig:Sensitivity} also displays the median sensitivity (i.e., limiting luminosity) as a function of the vertical distance, denoted as $D(z)$, which is the median value of all pixels within a narrow box positioned at $z$. 
The length of the box was chosen as the length of the major-axis of a given galaxy, while the width was set as $6\arcsec$. 
The source detection sensitivity throughout the \Chandra images varies, due to the different level of the diffuse X-ray emission related to the galaxy and the ICM, and the PSF degeneration at progressively larger off-axis angles. 

\subsection{Filtering nuclear and foreground sources} \label{subsec:result_OpticalCounterparts}

Prior to analyzing the spatial distribution of the X-ray sources (Section~\ref{sec:analysis}), we searched for and filtered any irrelevant sources. We identified AGN candidates by cross-matching our X-ray source catalog with the extended Virgo cluster catalog (EVCC) \citep{2014ApJS..215...22K}, 
which encompasses the spatial range of our Virgo LTG sample. Any X-ray source that matches within a $2''$ radius with the optical center was recognized as an AGN candidate and subsequently excluded from our analysis. The following eight galaxies host an AGN candidate: NGC\,4192, NGC\,4197, NGC\,4216, NGC\,4313, NGC\,4388, NGC\,4419, NGC\,4607 and VCC\,832. 
While not belonging to our LTG sample, VCC\,832 falls in the field of NGC\,4388.

We also searched for foreground stars by cross-matching with the \Gaia Data Release 3 \citep{2016A&A...595A...1G,2023A&A...674A...1G} point source catalogue. 
A foreground star is taken to have a parallax at least 5 times greater than its error, typically $\sim0.2 \ \mathrm{mas}$ at a  $G$-band magnitude of $18$. 
Only one source, found in the NGC\,4388 field, was identified as a probable foreground star. 
Thus we retain a total of 350 X-ray sources for the Virgo sample galaxies. 

Similarly, we identified the optical counterparts for the field LTGs. 
Eight galaxies were found to host an AGN candidate, namely NGC\,891, NGC\,3079, NGC\,3877, NGC\,4217, NGC\,4565, NGC\,4666, NGC\,5170 and NGC\,5746. Furthermore, ten X-ray sources were identified as foreground stars and excluded. 
This preserves 904 sources for the field sample.

\section{Spatial Distribution of the X-ray Sources} \label{sec:analysis}

\subsection{Off-disk Distribution of the X-ray Sources in Virgo LTGs} \label{subsec:result_SpatialDistribution}

\begin{figure*}
\includegraphics[width=0.5\textwidth]{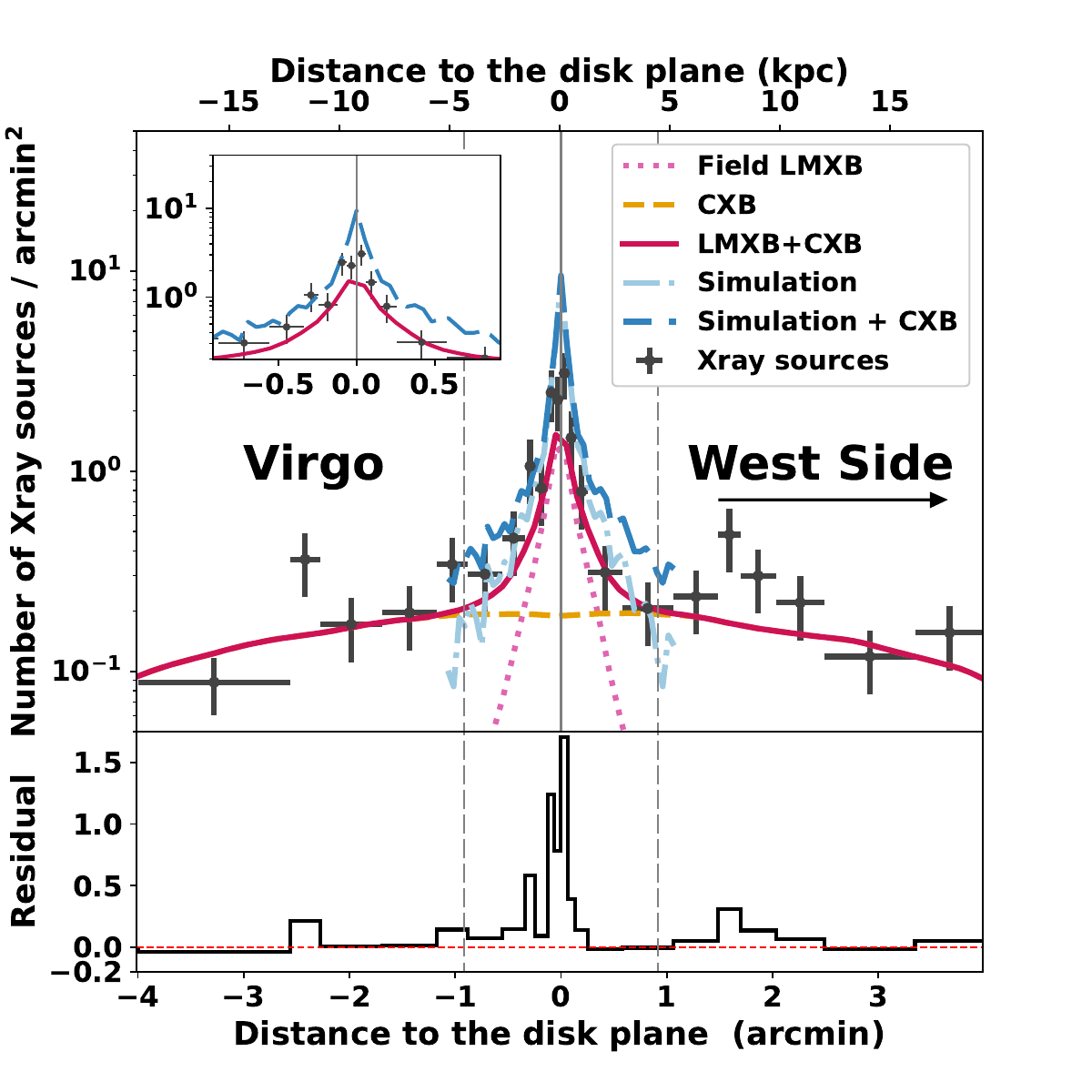}
\includegraphics[width=0.5\textwidth]{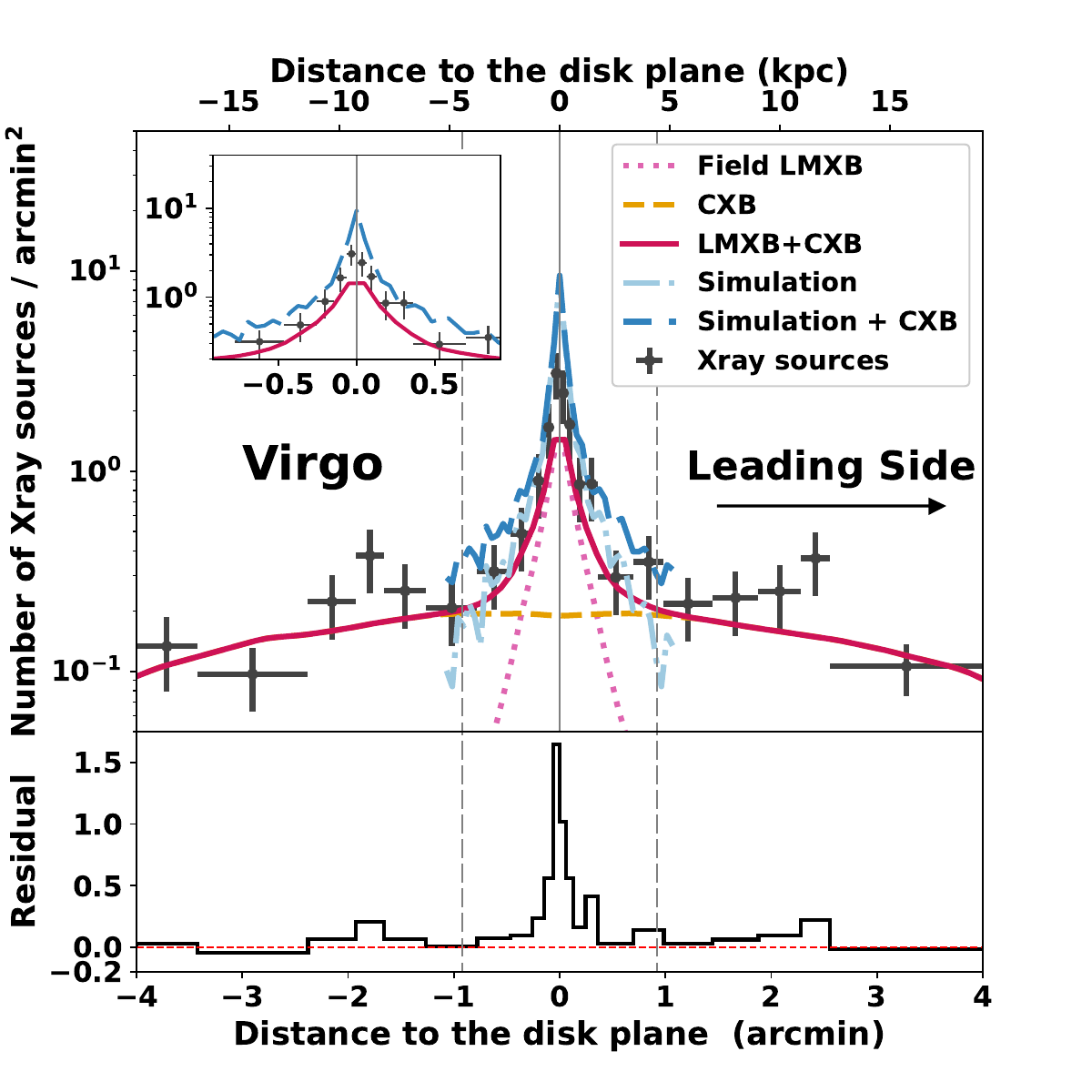}
\includegraphics[width=0.5\textwidth]{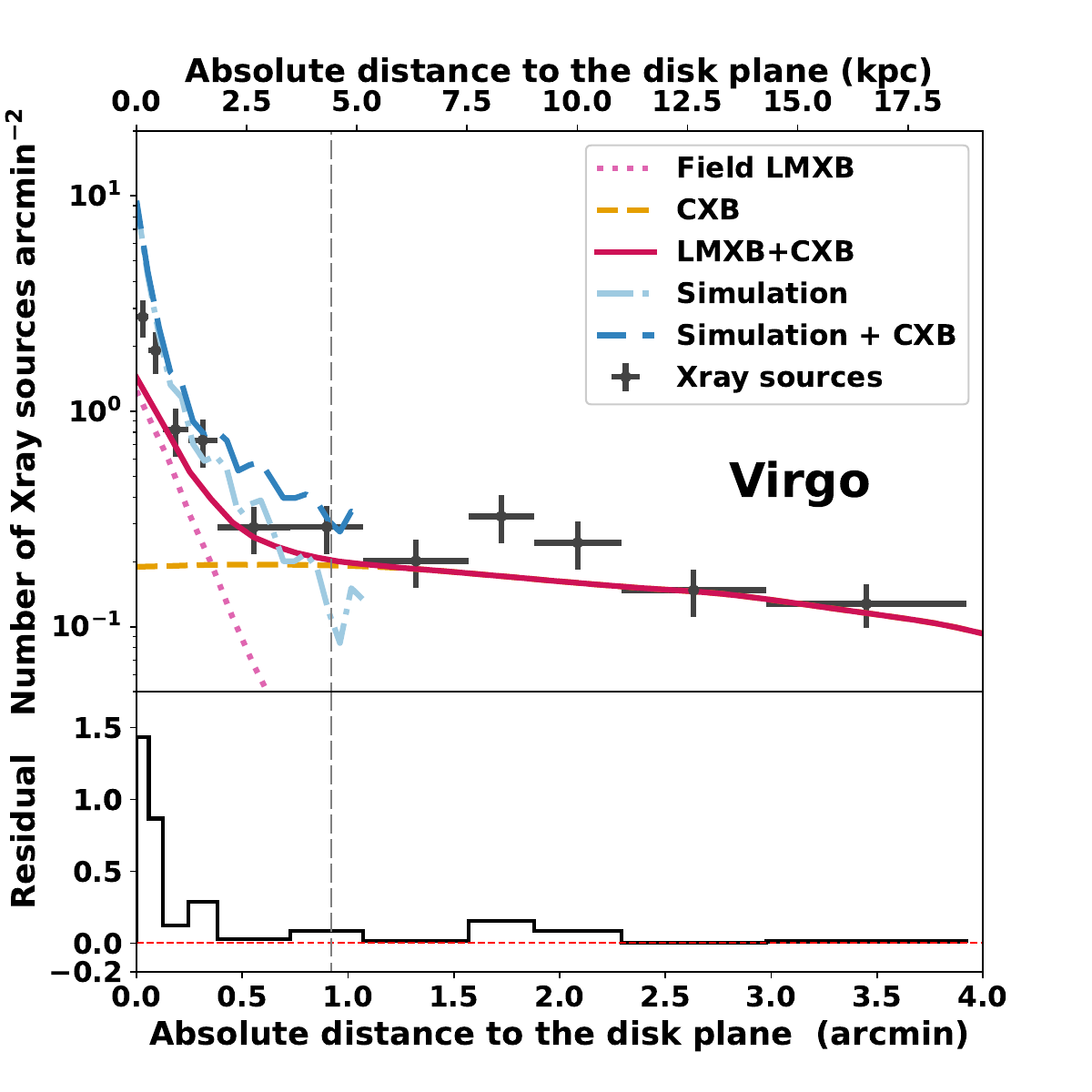}
\includegraphics[width=0.5\textwidth]{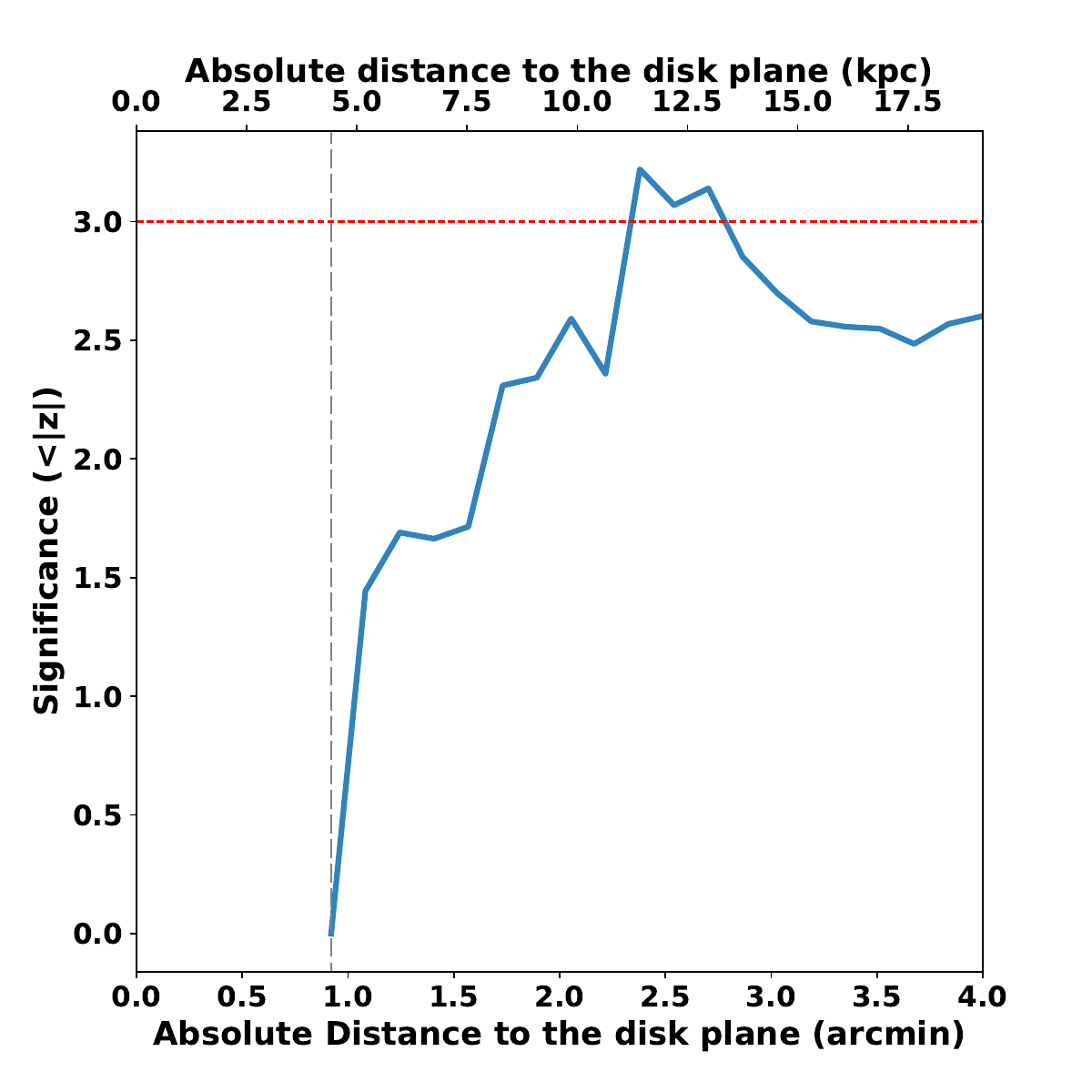}
\caption{The spatial distribution of X-ray point sources perpendicular to the disk plane of Virgo LTGs. {\it Upper left}: The $+z$ axis is aligned with the west side of the galaxies. The observed source profile (crosses) is adaptively binned to have at least 8 sources per bin. The orange dashed line, pink dotted line and red solid line represent the predicted CXB sources, field-LMXBs, and field-LMXB+CXB, respectively.
    The blue dash-dotted line represents the combined LMXB+HMXB distribution in a Milky Way-like galaxy predicted by the simulation of \citet{2008MNRAS.387..121Z}, whereas the blue long-dashed line includes the contribution from the CXB.
    The two vertical grey long-dashed lines mark the position of the mass-weighted mean semi-minor axis ($|z| = 0.92'$).   {\it Upper right}: Same as the upper left panel, but the $+z$ axis is aligned with the leading side of the host galaxies, determined by the ``maximum anisotropy cut" method \citep{2020MNRAS.497.4145T}. {\it Lower left}: Similar to the upper panels but combining both sides. The observed source profile is adaptively binned to to have at least 16 sources per bin. Within the range of $0.92'<|z|<2.5'$, an excess of off-disk X-ray sources with respect to the field-LMXB+CXB profile is evident. {\it Lower right}: The cumulative significance of excess over the predicted LMXB+CXB profile as a function of distance to the disk plane. 
\label{fig:Xraysrcprofile}}
\end{figure*}

To probe a putative excess of extraplanar X-ray sources, we construct a source surface density profile in the direction vertical to the major-axis of the disk.
For each galaxy, 
this profile includes X-ray sources detected within a box which has an equal length as the major-axis and a fixed half width of $4'\ (\sim19\ \mathrm{kpc}$ at the distance of Virgo). 
An example, for the case of NGC\,4216, is depicted by the light blue box in Figure \ref{fig:NGC4216_WISE_Chandra}. 
Due to the small number of sources detected in each individual galaxy, we stack all 19 Virgo LTGs to form a cumulative source surface density profile,  
as shown in the upper left panel of Figure \ref{fig:Xraysrcprofile}, where we align the west side with the $+z$ axis. 
Specifically, the mean X-ray source surface density is calculated by dividing the total number of sources detected within a certain range of distance, $z-{\Delta}z/2$ to $z+{\Delta}z/2$, by the total area, which is determined by multiplying ${\Delta}z$ by the total length of disk major-axis of all galaxies. The step ${\Delta}z$ is adaptively determined such that a minimum of 8 sources are enclosed, with a minimum value of $3\arcsec$.
This profile includes a total of 192 sources. 

It is expected that the X-ray source distribution consists of at least two primary components: (i) field-LMXBs that closely trace the stellar component of the host galaxies, and (ii) the CXB. 
We calculate the vertical distribution of these two components, 
taking into account the detection sensitivity described in Section \ref{subsec:datapreparation_X-ray}.
Specifically, the abundance of field-LMXBs is derived by integrating the X-ray luminosity function (XLF) of \citet[Eq.~2 therein] {2012A&A...546A..36Z}, down to the local sensitivity limit, as determined by $D(z)$. The normalization of the XLF is proportional to the enclosed stellar mass, which is measured according to the \WISE data\footnote{For regions far from the disk plane, where the $W1$ surface brightness is close to the sky level, the stellar mass surface density is set to be zero.}.  
We caution that this empirical XLF of field-LMXBs is derived from a sample of nearby ETGs \citep{2012A&A...546A..36Z}, and we have assumed that the same XLF can be applied to the field-LMXB population of LTGs. Similarly, we evaluate the CXB component by integrating the empirical $\log{N}-\log{S}$ relation of the CXB in the 0.5--8 keV band \citep[Eq.~3 therein]{2007ApJ...659...29K}, down to the local sensitivity limits. 

The predicted field-LMXBs and CXB profiles, and the sum of them, are plotted in Figure~\ref{fig:Xraysrcprofile}. We also depict the residual after subtracting the combined contribution of field-LMXB+CXB.
The field-LMXB component dominates $|z|\lesssim0.5'$, while CXB dominates when $|z|\gtrsim1'$, as expected. In the range of $0.4'\lesssim|z|\lesssim0.7'$, the predicted combined profile matches the observed X-ray surface density profile quite well. However, an overdensity occurs at $z\sim0'$, the disk mid-plane. This overdensity can be attributed to HMXBs, which are associated with star forming regions that distribute along the disk. However, the exact number of observable HMXBs is subject to a substantial uncertainty due to the nearly edge-on view, which introduces both source confusion and source obscuration by dusty cold gas in the disk. Therefore, we do not attempt to characterize the surface number density of HMXBs.

We now turn our focus to the off-disk sources. 
The edge of the disk region in the $z$ direction is defined as the stellar mass-weighted mean semi-minor axis ($|z|<0.92'$, corresponding to $\sim4.4\ \kpc$), as indicated by the vertical dashed lines in Figure \ref{fig:Xraysrcprofile}. 
The bottom left panel of Figure~\ref{fig:Xraysrcprofile} further displays the surface density profile as a function of the absolute off-disk distance (i.e., merging the two sides).
Between  $0.92'<|z|< 2.5'$, there appears an excess of X-ray surface density over the predicted field LMXB+CXB profile. The statistical significance of this excess (19.8 sources) is calculated according to the following equation:
\begin{equation}
    S = \sqrt{2}\ \mathrm{erf}^{-1} (1-Q(N,b)),
    \
    \label{eq:sig}
\end{equation}
where $\mathrm{erf}^{-1}$ is the inverse of the error function, $Q(N,b) = e^{-b}b^N / N!$ is the Poisson probability of detecting $N$ sources given an expected background of $b$ sources, which refers to the sum of the field-LMXB and CXB in this case. Thus $S$ indicates the confidence level for rejecting the null hypothesis that all $N$ sources are from the background. 
We find the excess between $0.92'<|z|< 2.5'$ with a significance of $3.3\sigma$. 
The bottom right panel of Figure~\ref{fig:Xraysrcprofile} shows the cumulative significance of the excess sources as a function of $|z|$ starting at $0.92'$. The median distance of all sources detected within this vertical range is $1.7'$ (8.1 kpc).
The significance drops to $2.9\sigma$ on the west side only and $2.3\sigma$ on the east side only. Thus, no statistical significant excess is present at either side alone.
We have verified that no single galaxy in the Virgo LTG sample dominates this off-disk excess. 
One galaxy, NGC\,4192, contributes $\sim 35\%$ of the total excess.
This can be attributed to NGC\,4192 having the highest SFR, second largest stellar mass and second longest exposure among the Virgo LTGs, as indicated in Table~\ref{table:VL_property}. 
The significance of the excess drops to $2.6\sigma$, if NGC\,4192 is removed from the Virgo sample.
We have neglected potential {\it cosmic variance} \citep{1992ApJ...396..430L} in the estimation of CXB sources. This cosmic variance is only $\sim5\%$ for the Virgo ACIS fields according to the estimation of \citet{2017ApJ...846..126H}, which is small compared to the Poisson error.
This is supported by the fact that the observed surface density profile at $|z| \gtrsim 2.5'$ ($\gtrsim$12 kpc) is fully consistent with the predicted CXB profile, which itself drops significantly with increasing $|z|$ due to the degrading source detection sensitivity.

We further examine the potential effect of ram pressure on the distribution of off-disk X-ray sources.  
This involves the identification of the ``leading side'' of an infalling galaxy, i.e., the side subject to ram pressure. 
We determine the leading side following the ``maximum anisotropy cut" method proposed by \citet{2020MNRAS.497.4145T}, which essentially finds a dividing line that maximizes the difference in SFR between the two halves of the disk, under the assumption that ram pressure would temporarily enhance star formation.
It turns out that 12 out of the 19 Virgo LTGs have their leading side consistent with the side closer to the center of M87, whereas the remaining 7 galaxies have their leading side facing away from M87.
We then align the leading sides with the $+z$ axis and reconstruct the stacked source profile, as shown in the upper right panel of Figure \ref{fig:Xraysrcprofile}. 
A halo excess of $3.3\sigma$ is still evident in this case, because the total number of detected sources preserves. 
Meanwhile, the excess at the leading and trailing  sides has a significance of $2.8\sigma$ and $2.4\sigma$, which 
again implies no statistically significant excess on either side alone.
The implication of this result is further discussed in Section \ref{subsec:dis:RPS}.

\subsection{Off-disk Distribution of the X-ray Sources in field LTGs}

\begin{figure*}
\includegraphics[width= 0.5\textwidth]{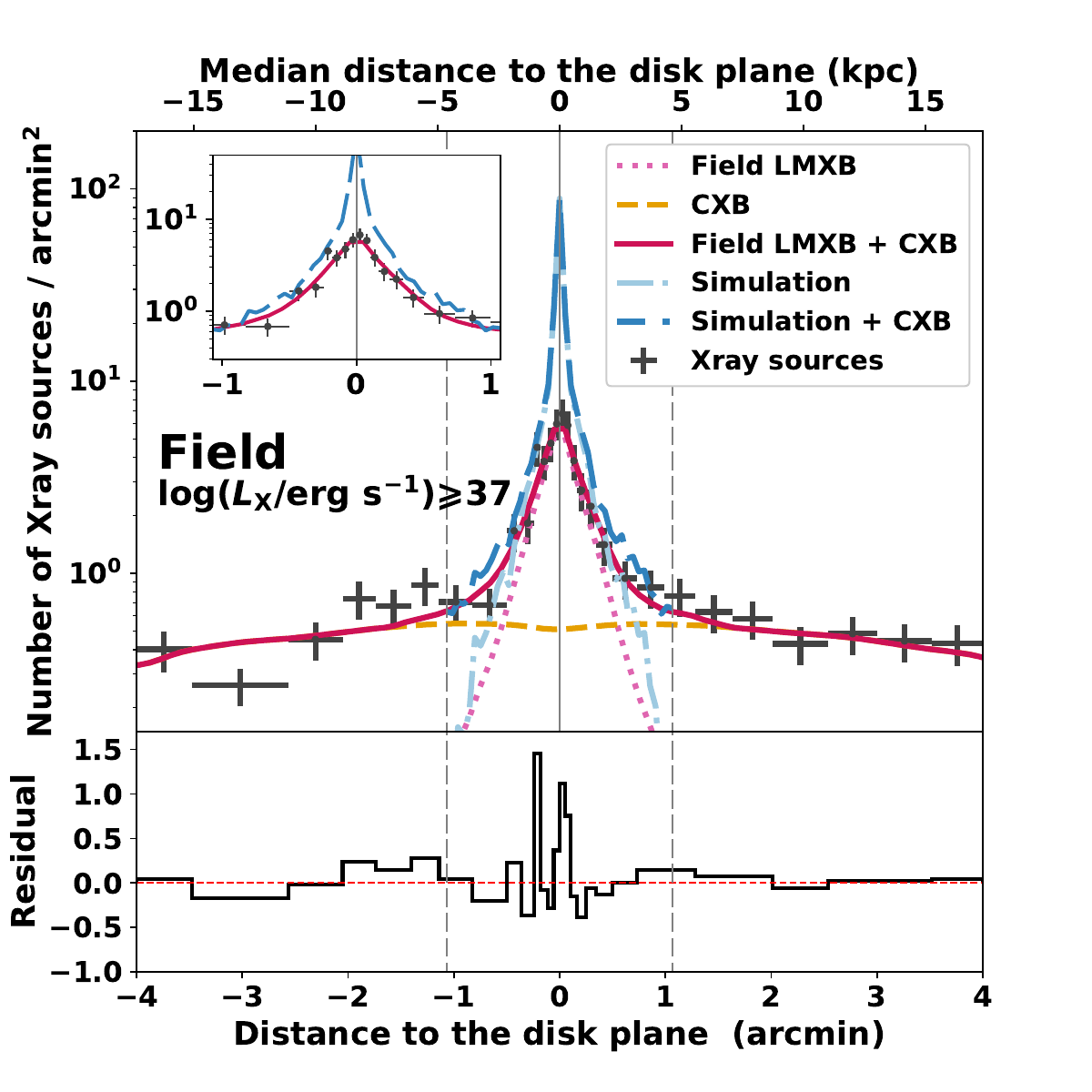}
\includegraphics[width= 0.5\textwidth]{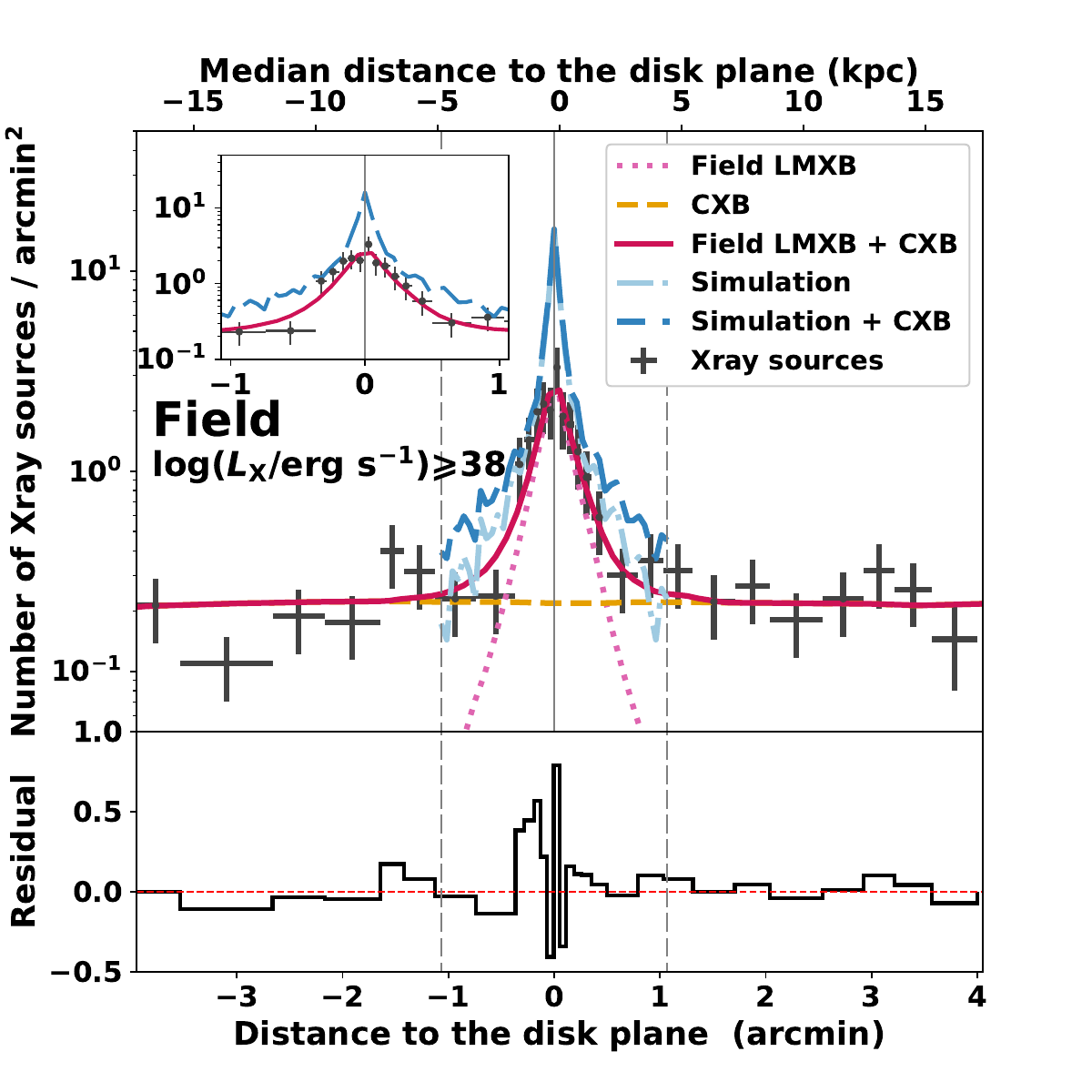}
\includegraphics[width= 0.5\textwidth]{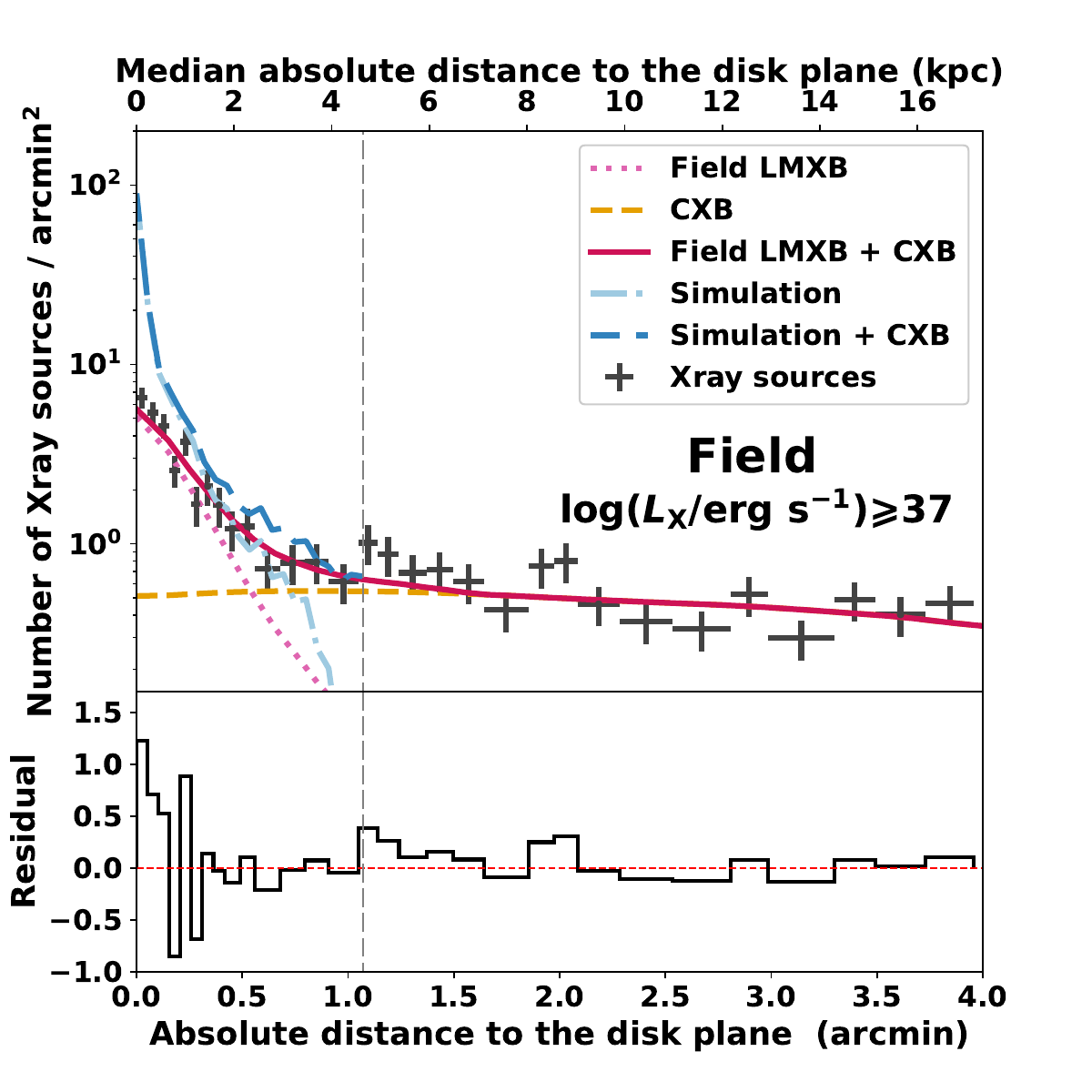}
\includegraphics[width= 0.5\textwidth]{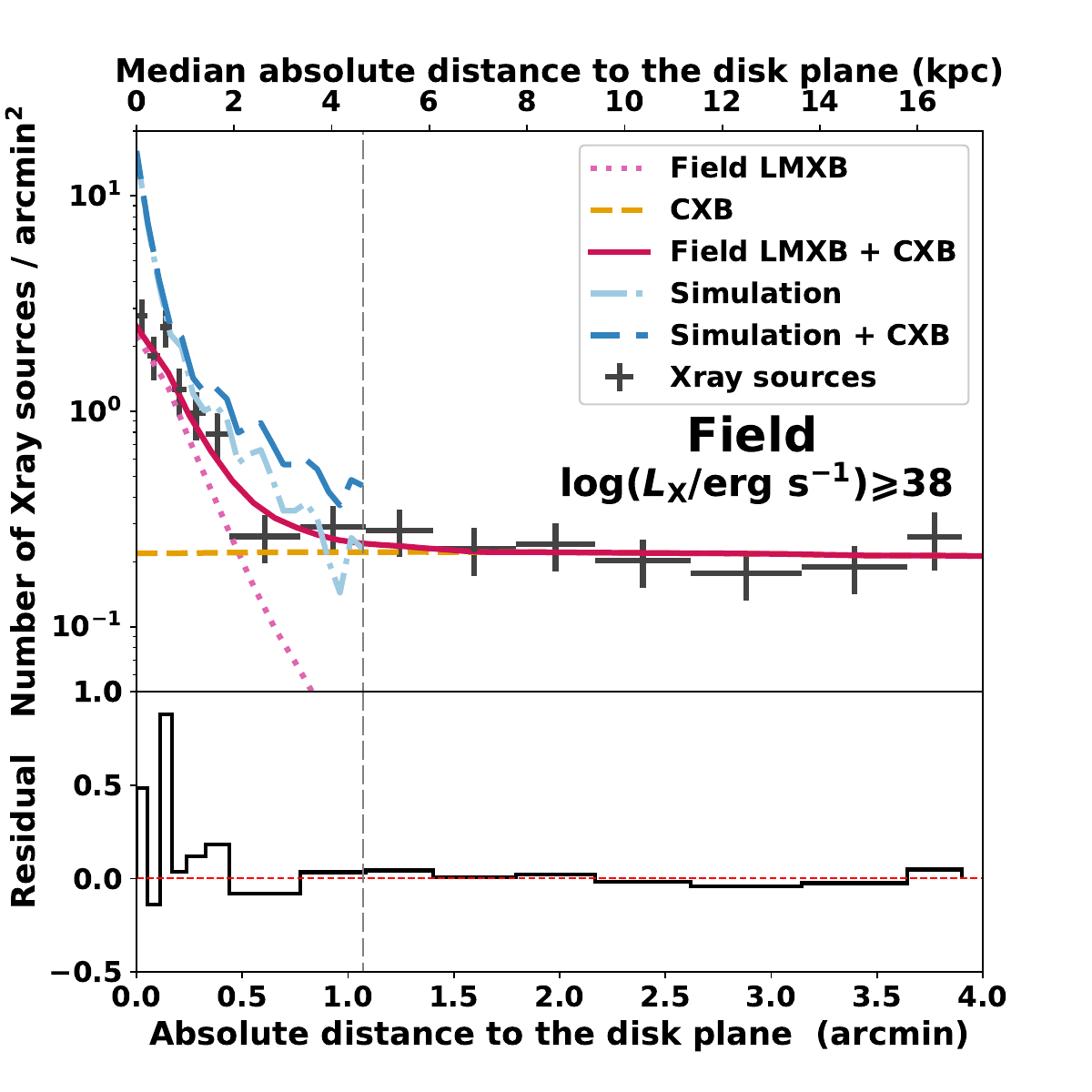}
\caption{The spatial distribution of X-ray point sources perpendicular to the disk plane of field LTGs. {\it Upper left}: The $+z$ axis is aligned with the near side of the galaxies. The observed source profile (crosses) is adaptively binned to have at least 8 sources per bin. The orange dashed line, pink dotted line and red solid line represent the predicted CXB sources, field-LMXBs, and field-LMXB+CXB, respectively.
    The blue dash-dotted line represents the combined LMXB+HMXB distribution in a Milky Way-like galaxy predicted by the simulation of \citet{2008MNRAS.387..121Z}, whereas the blue long-dashed line includes the contribution from the CXB.
    The two vertical grey long-dashed lines mark the position of the mass-weighted mean semi-minor axis ($|z| = 1.07'$).   {\it Upper right}: Same as the upper left panel, but only includes sources with an unabsorbed 0.5--8 keV luminosity above $10^{38}\rm~erg~s^{-1}$. {\it Lower left and lower right}: Similar to the upper panels but combining both sides. The observed source profiles (crosses) are adaptively binned to have at least 16 sources per bin.} %
    \label{fig:Field_profile}
\end{figure*}

Figure \ref{fig:Field_profile} shows the halo X-ray source distribution of the field galaxies. The $+z$ axis is aligned with the near-side of the galaxies, which is judged from the appearance of the in-disk dust lanes.  
The detection limit of the field LTGs are on-average $\sim5$ times deeper than the Virgo LTGs (Figure~\ref{fig:Sensitivity}).
Therefore, we consider two sets of vertical source surface density profiles. 
The first set includes all sources detected within a rectangle, again with a length equaling to the major-axis of the disk and a fixed  width of $8\arcmin$. 
The second set includes only the bright sources in this rectangle, which have an inferred 0.5--8 keV luminosity $\logLx > 38$, to be more consistent with the Virgo sample.
The two sets of profiles, which include 619 and 258 sources, respectively, are shown in the
left and right columns of Figure~\ref{fig:Field_profile}. In this figure, the upper rows display the profiles on both sides whereas the lower rows display the profiles merging the two sides.
We also plot the predicted field-LMXB and CXB components for these field LTGs, accounting for the respective detection limits. Likewise, the residual after subtracting the field-LMXB+CXB contribution is shown.

Similar to the case of Virgo LTGs, an overdensity of sources is seen near the mid-plane, which can again be understood as the unaccounted contribution by HMXBs.
We evaluate the potential excess of off-disk sources within the range of $1.07'<|z|<2.5'$, with 1.07$\arcmin$ (corresponding to 4.6 kpc at the median distance of the field LTGs) being the mass-weighted mean semi-minor axis of the field LTGs. 
Using Eq.~\ref{eq:sig}, an excess with a significance of $3.0\sigma$ is found  including both sides of the profile constructed at the full detection limit. 
However, no significant excess ($2.0\sigma$) is found in the profile of bright sources (i.e., above the limit of $\logLx=38$).
We also verify that no single galaxy dominates the number of excess sources. 

We summarize the information about the off-disk excess in Table \ref{table:excess_sig}.

\begin{deluxetable*}{cccccccc}
\tablenum{2}
\tablecaption{Excess Off-disk X-ray Sources in Different Groups}
\tablewidth{0pt}
\tablehead{
\colhead{Group}& \colhead{Detection Limit} &       \colhead{Range} &  \colhead{$N_{\rm obs}$} &  \colhead{$N_{\rm LMXB}$} &   \colhead{$N_{\rm CXB}$} &  \colhead{$N_{\rm excess}$} &  \colhead{Significance} 
}
\decimalcolnumbers
\startdata
 &       &   $0.92'\atob2.5'$ &    34 &     0.1 &   21.8 &     12.1 &  $2.9\sigma$ \\
  Virgo, west half aligned          &       $\sim10^{38}$ &  $-2.5'\atob-0.92'$ &  30 &     0.2 &   22.1 &      7.7 &  $2.3\sigma$ \\
            &                 &              Total &  64 &     0.3 &   44.0 &     19.8 &  $3.3\sigma$ \\
\hline
 &       &   $0.92'\atob2.5'$ &   33 &     0.2 &   21.7 &     11.2 &  $2.8\sigma$ \\
  Virgo, leading half aligned       &      $\sim10^{38}$             &  $-2.5'\atob-0.92'$ &  31 &     0.1 &   22.3 &      8.6 &  $2.4\sigma$ \\
     &            &              Total &      64 &     0.3 &   44.0 &     19.8 &  $3.3\sigma$ \\
\hline
 &        &   $1.07'\atob2.5'$ &   73 &     1.9 &   64.8 &      6.3 &  $2.1\sigma$ \\
Field, all            &     $\sim10^{37}$             &  $-2.5'\atob-1.07'$ &   84 &     1.7 &   62.3 &     20.0 &  $3.0\sigma$ \\
            &                 &              Total &    157 &     3.6 &  127.1 &     26.3 &  $3.0\sigma$ \\
\hline
 &       &   $1.07'\atob2.5'$ &  29 &     0.5 &   27.9 &      0.6 &  $1.8\sigma$ \\
Field, bright     &      $\sim10^{38}$      &  $-2.5'\atob-1.07'$ &        30 &     0.4 &   27.4 &      2.2 &  $1.8\sigma$ \\
            &                 &              Total &      59 &     0.9 &   55.3 &      2.8 &  $2.0\sigma$ 
\enddata
\label{table:excess_sig}
\tablecomments{(1) Groups of off-disk X-ray sources. (2) Approximate detection limit, in units of erg~s$^{-1}$. (3) The vertical distance range. (4) Number of observed X-ray sources. (5) Predicted number of field LMXBs. (6) Predicted number of CXB sources. (7) Number of excess sources. (8) Significance of the excess.  }
\end{deluxetable*}

\section{Discussion}
\label{sec:discussion}

In the previous section, we have studied the spatial distribution of X-ray sources in 19 Virgo LTGs and 12 field LTGs, the disk of which has a high inclination angle that allows for a clear view of off-disk sources. 
We find a significant ($3.3\sigma$) excess of 19.8 sources in the halo (within a mean vertical distance of 4.4--12.0 kpc from the disk) of the Virgo LTGs with respect to the expected total number of CXB and field-LMXBs sources, down to a detection limit of $\logLx \gtrsim 38$. 
On the other hand, in the field LTGs, no significant excess is found at a similar detection limit. 
Only when the full source detection limit ($\logLx \gtrsim 37$) is considered for the field LTGs, an extraplanar excess becomes apparent, with a significance of $3.0\sigma$.
This contrast is unlikely caused solely by the difference in the number of galaxies involved, because both the total stellar mass and total SFR, the two main parameters presumably determining the number of X-ray binaries, are actually $\sim2.5$ higher for the field LTGs than the Virgo LTGs (Table~\ref{table:VL_property}).
This suggests that the excess sources around the Virgo LTGs is at least partly related to the cluster environment. 
Below we discuss possible origins of this excess, after a close comparison with the halo excess previously found in Virgo ETGs.

\subsection{Comparison with early-type galaxies}
\label{subsec:ETG}
Using {\it Chandra} observations of 80 Virgo ETGs in the stellar mass range of $\sim10^9-10^{11}\rm~M_\odot$,  \citet{2017ApJ...846..126H} detected an excess of X-ray sources beyond three times the median effective radius (2.4--19.2 kpc) at a significance of $3.5\sigma$. Similarly, no significant excess was found by these authors in their field analogs, which led them to suggest a cluster-related origin of the excess.  
The detection limit of the Virgo ETG and LTG samples are quite similar ($\logLx\gtrsim38$).
The ETG sample is found to have $N_{\rm excess}=116.0$ sources, which gives an expected number of 27.6 excess sources for the LTG sample if we simply scale the number of galaxies.
If instead we scale the total stellar mass ($9\times10^{11}\ M_\odot$ in the ETGs vs. $3\times10^{11}\ M_\odot$ in the LTGs), the expected excess would be 38.7 sources in the LTGs.
These values are 1.4--2 times higher than the actually observed number of excess sources (19.8; Table~\ref{table:excess_sig}). 
On the other hand, 
the projected sky area within which the ETG halo excess is found (an annulus between $0.5'<R<4'$, the projected area of which is 49.5 arcmin$^2$; \citealp{2017ApJ...846..126H}) is $\sim$3.7 times higher than for the LTG halo excess (two boxes at $0.92'<|z|<2.5'$, for a mean major-axis of $4.2'$, the projected area is $13.3$ arcmin$^2$). 
Thus, it is expected that 31.1 excess sources will be detected in the LTG sample when scaling the projected area, comparable to the observed number.

Even though both samples are Virgo member galaxies, the origin of the excess in the LTGs and ETGs may be different due to their different properties and varied interactions with the cluster environment. 
Below we consider several potential origins for the excess X-ray sources around the LTGs, which include GC-LMXBs, SN-kicked X-ray binaries, intra-cluster X-ray binaries, wandering massive black holes and X-ray binaries induced by ram-pressure stripping.

\subsection{GC-LMXBs} \label{subsec:GC}

The first candidate for the off-disk excess sources is GC-LMXBs, i.e., LMXBs residing in globular clusters, which are often found in both LTGs and ETGs and can have a spatial distribution more extended than the bulk of the starlight \citep{2006ARA&A..44..323F}.
However, previous researches on ETGs find that GC-LMXBs only have a limited contribution to the excess sources. For example, in the Virgo cluster, \citet{2017ApJ...846..126H} found that $\sim 30\%$ excess sources with $\logLx>38$ are GC-LMXBs. For the Fornax cluster, \citet{2019ApJ...876...53J} estimated that $\lesssim20$ out of $74$ excess sources around the central giant elliptical galaxy NGC\,1399 could be associated with faint, undetected GCs.   
In addition, \citet{2023ApJ...953..126H} noticed that LTGs have on-averge fewer GC-LMXBs than ETGs. In particular, $1.2\%$ GCs in LTGs contain GC-LMXBs, while for ETGs, this rate rises to $4\%\textup{--}10\%$. 

A rough estimate of the number of GC-LMXBs can be derived as follows. According to a study on several face-on spirals, the average number of GCs per LTG is $198$ inside the isophotal ellipse that traces the $K_s\approx20\ \mathrm{mag\ arcsec^{-2}}$ galactic surface brightness \citep[Table 5 therein]{2023ApJ...953..126H}. 
Assuming that those GCs follow the spatial distribution of known GCs in the Milky Way (a total of 157), we expect $16\%$ (25/157) to fall in the range of $4.4\ \kpc <|z|<12.0\ \kpc$ \citep{1996AJ....112.1487H}, i.e., the Virgo LTG off-disk range where excess sources are found. 
Again from \citet{2023ApJ...953..126H}, $1.2\%$ GCs contains LMXBs and $24\%$ (6/25) of these are brighter than our median detection limit of $\logLx\gtrsim38.2$.
As a result, for the 19 Virgo LTGs, we estimate that GC-LMXBs contribute only $\sim9\% \ (1.7/19.8)$ of the excess sources.

\subsection{SN-kicked XRBs} \label{subsec:SN}

The second candidate is SN-kicked LMXBs.
A compact object (a neutron star or a stellar-mass black hole) born after a core-collapsed supernova explosion can receive a strong kick. 
If this compact object were part of a binary system, 
the binary may survive depending on the resultant kick velocity, and later, further depending on the gravitational potential of the host galaxy, escape into the galactic halo and manifest itself as an XRB, most likely an LMXB due to a sufficiently long lifetime \citep[e.g.][]{1995MNRAS.274..461B,2008MNRAS.387..121Z}. 
This effect can result in a flattened spatial distribution  of X-ray point sources \citep[e.g.][]{2013A&A...556A...9Z}.  

\citet{2008MNRAS.387..121Z} simulated the spatial distribution of XRBs (including both LMXBs and HMXBs) in a Milky Way-like galaxy, explicitly including the effect of supernova kick. We plot their simulated vertical source surface density distribution in Figure \ref{fig:Xraysrcprofile} and Figure \ref{fig:Field_profile}  with blue lines, which assumes a perfectly edge-on view. 
It is noteworthy that this synthetic profile does not take into account dust obscuration or source confusion.
This may explain the apparent excess with respect to the observed source surface density at $|z| \sim 0$.
Nevertheless, the synthetic profile is significantly flattened 
at $|z|\sim1'$, i.e., near the edge of the disk, which is consistent with the qualitative behavior of the observed source profiles, in particular that of the field LTGs of the full detection limit.
Unfortunately, \citet{2008MNRAS.387..121Z} only presented the synthetic profile out to a vertical distance of 5 kpc, preventing a comparison with the observed profile at larger heights. 
Another limitation of the simulations is that only a fiducial Milky Way-like SFR of $ \sim 2\ \Msunyr$ and a fixed stellar mass ($\sim10^{11}\rm~M_\odot$) are assumed (Z. Zuo, private communication), while our LTG samples span a substantial range of SFR and stellar mass (Table~\ref{table:VL_property}).

Nevertheless, we suggest that SN-kicked XRBs may account for a good fraction, if not the majority, of the observed off-disk excess sources.
First, such a population is a generic result of normal stellar evolution in a typical galaxy. SN-kicked XRBs have been evidenced around both low-to-intermediate mass ETGs of Virgo \citep{2017ApJ...846..126H} and the brightest cluster galaxy of Fornax \citep{2019ApJ...876...53J}.  
Second, this helps to explain the detection/non-detection of excess sources (at the limit of $\logLx\gtrsim38$) in the Virgo/field LTGs, despite the former having substantially smaller stellar mass and lower current SFR (this latter factor could be less relevant if the SFR was higher in the past). 
While a smaller galaxy with a lower SFR is expected to produce fewer SN-kicked binaries, its shallower gravitational potential would allow the binaries to ballistically travel to a greater height, as demonstrated by the simulation of \citet{2008MNRAS.387..121Z}.
Further simulations sampling a large range of SFR and stellar mass would be desired for a more quantitative comparison with the observation.

\subsection{Intra-cluster X-ray sources}\label{subsec:ICL_wandering}

The third candidate is ICXs, originally proposed to exist in both Virgo \citep{2017ApJ...846..126H} and Fornax \citep{2019ApJ...876...53J}.
In a broad context, ICXs may include LMXBs, HMXBs and wandering MBHs. 
By definition, ICXs are currently gravitationally unbound to any particular galaxy, although they could have originated from some certain galaxy, as the case of the optical diffuse ICL \citep{2022NatAs...6..308M}. 
It is believed that the ICL is formed by the stripped and expelled material from cluster galaxies due to their complex dynamical processes, including frequent tidal encounters and mergers \citep{2021Galax...9...60C}.
Virgo is among the first clusters in which the presence of ICL is established \citep{2005ApJ...631L..41M}. 
The red color of the ICL \citep[e.g.][]{2017ApJ...834...16M} implies predominantly old stellar populations, which are naturally a reservoir of LMXBs (ICL-LMXBs). 
Strictly determining whether an X-ray source found off-disk is bound to the host galaxy requires kinematic information, which is currently unavailable even for the nearby galaxies studied here.
Nevertheless, it is still plausible that some of the off-disk excess sources around the Virgo LTGs are truly ICXs. 

Following \citet{2017ApJ...846..126H}, we estimate the number of ICL-LMXBs as $N({\rm ICL-LMXB}) = \epsilon f_{\rm FoV} f_{\rm ICL} f_{\rm stars} M_{200}$, where $M_{200}=1.05\pm0.02\times10^{14}\ \mathrm{M_\odot}$ is the viral mass of Virgo \citep{2017MNRAS.469.1476S}, $f_{\rm stars}=0.05$ is the fractional stellar mass according to the empirical measurements for clusters that are as massive as Virgo \citep{2007ApJ...666..147G},  $f_{\rm ICL}\sim 10\%$ is the fraction of total stellar mass contained in the ICL component \citep{2017ApJ...834...16M} and  $f_{\rm FoV} = 0.2\%$ is the fractional coverage of the excess region with respect to the projected area of the Virgo cluster within the viral radius of 974 kpc ($3.38^\circ$). 
The abundance of LMXBs above a mean detection limit of $\logLx\geqslant38$  ($\epsilon=2.5$ per $10^{10}\ \mathrm{M_\odot}$), is derived based on the XLF of \citet{2012A&A...546A..36Z}, which is also used for estimating the field-LMXBs. 
However, this results in only $0.3$ ICL-LMXBs, far less than the $\sim10$ ICL-LMXBs estimated around the Virgo ETGs \citep{2017ApJ...846..126H}.
This may be attributed to both a smaller galaxy sample and a smaller projected area from which the excess sources are probed.

In addition to old stars, the ICL may also contain freshly born stellar populations, which, for instance, arise from gas stripped from member galaxies and subsequently cool and collapse to form new stars in the intra-cluster space. HMXBs may be formed via such a channel and manifest as part of the ICX for a relatively short timescale ($\sim10^7$ yr).  
This possibility is further discussed in Section~\ref{subsec:dis:RPS}.

Moreover, MBHs wandering in the intra-clsuter space and accreting from the ICM may also shine in X-rays and become part of the ICXs. 
Such MBHs can be in place as the relic of tidally stripped satellite galaxies \citep[e.g.][]{2009ApJ...699.1690M,2010ApJ...721L.148B}, or due to gravitational recoil after the merger of two MBHs \citep[e.g.][]{2003ApJ...582..559V,2009MNRAS.395..781O,2013CQGra..30x4007S}. 
Due to the rich history of galaxy interaction and merger in a cluster like Virgo, there could be numerous wandering MBHs. However, an MBH fed directly by the hot and tenuous ICM is expected to be a relatively faint X-ray source. \citet{2017ApJ...846..126H} estimated that there should be less than 10 wandering MBHs among their $\sim120$ excess X-ray sources around the Virgo ETGs. Assuming a similar ICM condition around the LTGs, this translates to only few, and possibly none, MBHs among the off-disk excess sources.

Therefore, we conclude that ICXs are unlikely to be the main source of the excess found around the Virgo LTGs.

\subsection{Are there X-ray sources related to ram presure stripping?} \label{subsec:dis:RPS}

When a galaxy falls into a cluster, the ICM exerts ram pressure on it \citep{2022A&ARv..30....3B}.  The ram pressure can strip away gas belonging to the galaxy and eventually quench the process of star formation. 
On a shorter timescale, ram pressure can also compress gas and temporarily enhance star formation. For example, according to the EAGLE simulation, \citet{2020MNRAS.497.4145T} found an enhancement of ISM pressure and SFR on the leading side of the infalling galaxy. 
This raises the interesting possibility of ram pressure-induced halo X-ray sources. 
On the one hand, the enhanced SFR on the leading side might result in some SN-kicked XRBs. Given a sufficiently large kick velocity (several hundred $\rm~km~s^{-1}$), some of these XRBs, even an HMXB with a lifetime of $\sim10^7$ yr, might travel a distance up to a few kpc into the halo. This would preferentially lead to the detection of X-ray sources on the leading side of the galaxy. 
On the other hand, as mentioned in Section~\ref{subsec:ICL_wandering}, star formation might take place in the ram pressure stripping tail, which would result in the detection of X-ray sources on the trailing side of the galaxy.
Our exercise of aligning the Virgo LTGs according to their leading side (Figure~\ref{fig:Xraysrcprofile}) shows that the marginal excess on either side of the galaxy is rather comparable. 
This implies that both processes may contribute to some extent.
The role of ram pressure stripping, which uniquely applies to the cluster environment, also helps to explain the fact that the off-disk excess is more significant in the Virgo LTGs.

\section{Summary} \label{sec:Summary}

We have conducted a systematic search for X-ray point sources in the halo of 19 nearly edge-on Virgo LTGs and 12 field analogs, based on archival \textit{Chandra}  observations. 
Our main findings are as follows:

\begin{itemize}
\item We detect $64$ extraplanar sources at a detection limit of $\logLx\gtrsim 38$.  
Statistically accounting for the contributions of CXB and field-LMXB sources, we identify an excess of $19.8$ extraplanar sources within a range of $0.92'<|z|<2.5'$, at a significance level of $3.3\sigma$. 

\item On the other hand, no significant excess is observed in the cumulative source vertical profile of the 12 field LTGs at a similar detection limit, which implies that the excess sources around the Virgo LTGs could be at least partly related to the cluster environment. 

\item We find that GC-LMXBs or ICXs can have only a minor contribution to the observed excess,
whereas SN-kicked LMXBs might contribute a good fraction of the excess. In particular, even though the Virgo LTGs might have produced fewer SN-kicked LMXBs due to their on-average smaller stellar mass, their shallower gravitational potential would allow the SN-kicked LMXBs to travel a greater distance into the halo.

\item HMXBs recently formed due to the effect of ram pressure stripping are also likely to have a substantial contribution to the observed excess.  However, we do not find a clear enhancement on the leading side of Virgo LTGs, which might be understood as a balance of the SFR enhanced on the leading side and the SFR that take place in the ram pressure stripping tail. 

\end{itemize}

Wide-field X-ray observations, such as those afforded by the on-going eROSITA survey, would be useful to probe relatively bright X-ray sources in and around a large number of nearby galaxies, in both the field and cluster environments, as demonstrated by the recent X-ray census of active galactic nuclei in Virgo based on the first all-sky survey data of eROSITA \citep{2024arXiv240207275H}.

\begin{acknowledgments}

Z.H. and Z.L. acknowledge support by the National Natural Science Foundation of China (grant 12225302) and the National Key Research and Development Program of China (NO.2022YFF0503402). 
M.H. is supported by the National Natural Science Foundation of China (12203001) and the fellowship of China National Postdoctoral Program for Innovation Talents (grant BX2021016). 
We thank Lin He, Tom Jarrett, Zhanhao Zhao and Zhaoyu Zuo for helpful discussions.

This paper employs a list of \Chandra datasets, obtained by the Chandra X-ray Observatory, contained in \dataset[DOI: 10.25574/cdc.221]{https://doi.org/10.25574/cdc.221}.

This research made use of \texttt{photutils}, an \texttt{astropy} package for
detection and photometry of astronomical sources \citep{larry_bradley_2021_5796924}.
\end{acknowledgments}

\bibliography{VirgoLTG}{}
\bibliographystyle{aasjournal}

\end{document}